# From atomic physics, to upper-atmospheric chemistry, to cosmology:
## A "laser photometric ratio star" to calibrate telescopes at major observatories

Justin E. Albert ,*[a] Dmitry Budker ,[b,c,d] and H. R. Sadeghpour [e]

[a] Department of Physics and Astronomy, University of Victoria, Victoria, British Columbia V8W 3P6, Canada
[b] Johannes Gutenberg-Universität Mainz, 55128 Mainz, Germany
[c] Helmholtz-Institut, GSI Helmholtzzentrum für Schwerionenforschung, 55128 Mainz, Germany
[d] Department of Physics, University of California, Berkeley, California 94720, USA
[e] ITAMP, Center for Astrophysics | Harvard & Smithsonian, Cambridge, Massachusetts 02138, USA

**Correspondence**
Justin E. Albert, Department of Physics and Astronomy, University of Victoria, Victoria, British Columbia V8W 3P6, Canada.
Email: jalbert@uvic.ca

This Research Highlight showcases the two Research Papers entitled, *A precise photometric ratio via laser excitation of the sodium layer – I. One-photon excitation using 342.78 nm light*, https://doi.org/10.1093/mnras/stab1621 and *A precise photometric ratio via laser excitation of the sodium layer – II. Two-photon excitation using lasers detuned from 589.16 nm and 819.71 nm resonances*, https://doi.org/10.1093/mnras/stab1619.

*"The instruments of darkness tell us truths"*
Macbeth, Act I scene 3

**The nature of dark energy —** that enigmatic aspect of spacetime that apparently comprises over two-thirds of the total content of the Universe — is still elusive, more than twenty years after its discovery [1,2]. There has been outstanding progress in astronomy over these two decades, in particular in the development of observatories that will be able to image broader swaths of our Universe at unprecendented depths (such as the newly-launched James Webb Space Telescope [JWST]; and the also-soon-to-be-operational Vera C. Rubin Observatory in Chile — which will perform a full-southern-sky survey this decade that will be known as the "Legacy Survey of Space and Time" [LSST]) [3,4]. However, the equally-important problem of the *precision* of measurements of the brightness of astronomical objects, i.e. the uncertainties on astronomical magnitudes, remains a limiting factor on measurements of dark energy, and of the accelerating expansion of the Universe. In particular, the uncertainties on the magnitudes of "supernovae of type Ia" (abbreviated as "SNe Ia") — i.e. supernovae that are caused by the collapse of a white dwarf star — are presently the primary limitation on measurements of the nature of dark energy [5]. While multiple new observatory surveys this decade such as LSST, and those to be performed by JWST, will vastly increase the *number* of SNe Ia that are observed and measured, the *precision* of the measurements of SNe Ia magnitudes (and, in particular, the precision on measurements of *ratios* of SNe Ia magnitudes measured in visible bands, vs. magnitudes of the same SNe Ia when measured in near-infrared bands [6]) will remain as the main limitation on measurements of dark energy, in the absence of additional novel technology for the calibration of astronomical magnitudes as a function of color, at unprecedented levels of precision [7].

Two critical related technologies, those of laser guide stars (LGS) that are already used in astronomy (for example, an image of an LGS at the William Herschel Telescope on La Palma is shown in Figure 1), and modifications of existing LGS in order to utilize additional properties of sodium atoms in the Earth's upper atmosphere, will be key to solving this persistent problem of "spectrophotometric precision" in astronomy, and to improve our understanding of dark energy. LGS have been used since the 1990s to solve an entirely separate problem in astronomy: the *distortion* of the shapes of images of astronomical sources due to turbulence in the Earth's atmosphere [9]. Sodium LGS work by using a ground-based laser, at a wavelength of approximately 589 nm, to excite the $D_2$ resonance of sodium atoms in the mesosphere. Neutral sodium atoms are a minor component of Earth's upper atmosphere, which are predominantly located in the upper mesosphere between about 80 km and 105 km above sea level [10]. These sodium atoms originate primarily from the ablation of meteors [11]. Although atomic sodium is, overall, a minor component of our atmosphere, its large optical cross-section makes it the most favorable atmospheric component for optical excitation [12].

In general, LGS don't help at all with the problem of precisely measuring the magnitudes of astronomical sources. Moreover, the problem of precisely measuring magnitudes — *unlike* the problem of eliminating atmospheric distortion — is unfortunately *not* "simply solved" by moving one's telescope above the Earth's atmosphere into space [13]. However, by exciting certain alternative resonances (i.e., instead of the $D_2$ resonance) of the upper-atmospheric sodium atoms, one can create a mandatory "cascade" of sodium de-excitations, that will then create a two-color artificial star with a precisely 1:1 ratio of yellow (589 nm) to near-infrared (820 nm) photons that are produced in the upper atmosphere. Since the precision of that 1:1 photon production ratio will be known to better than a part in $10^4$ (i.e., better than 0.01%), such a "laser photometric ratio star" (LPRS) would allow the calibration of astronomical magnitudes

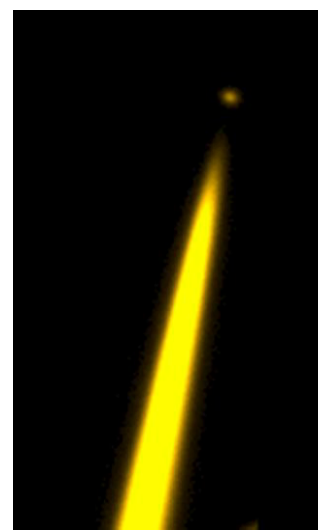

Figure 1: A pseudocolor image of a laser guide star (LGS) beacon located near the William Herschel Telescope on La Palma, Spain, taken using a small portable telescope placed 3 m from the laser launch site, is shown [8].

In visible light, a laser photometric ratio star (LPRS) would be expected to appear fairly similar to this image.

However, an LPRS itself (analogous to the roughly circular spot — above the long tail of Rayleigh-scattered light below it) would additionally emit a precisely equal amount of near-infrared light (at an 820 nm wavelength), as 589 nm light (the wavelength shown in this image).

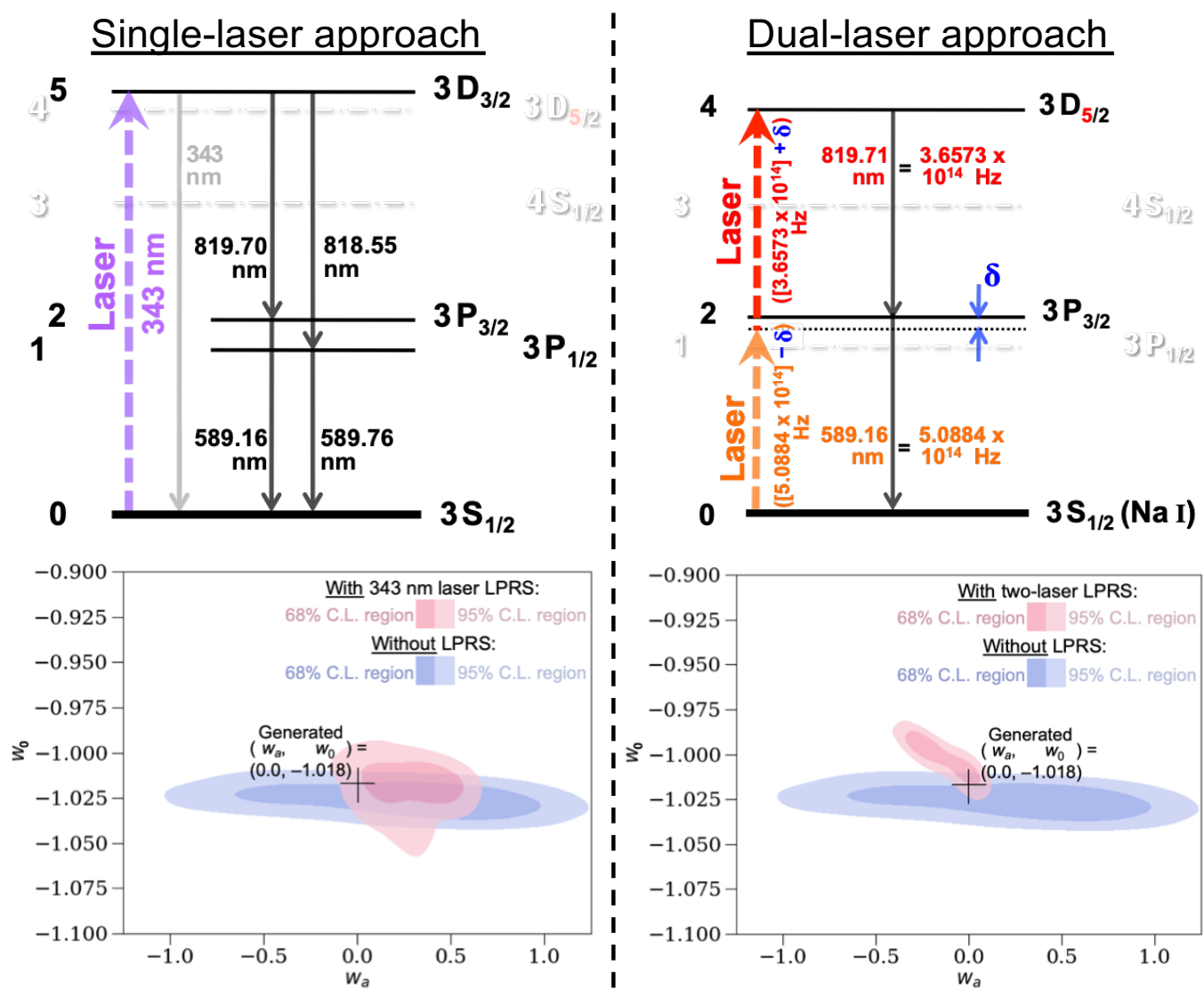

Figure 2: The two approaches for generating a laser photometric ratio star (LPRS), and their respective expected impacts on measurements of dark energy cosmological parameters, are shown in this figure. The diagrams at top left and top right show atomic levels (not to scale) for neutral sodium atoms (Na I) within the Earth's upper atmosphere, starting from their ground state ($3S_{1/2}$); and showing the atomic states reached via excitation by light from one, or from two, ground-based lasers. In both of the atomic level diagrams, the allowed and the laser-excited atomic states are shown as solid black lines; the "forbidden" 343 nm electric quadrupole de-excitation in the upper left diagram is shown as a gray solid downward-pointing arrow; while "ghost" levels, that are entirely inaccessible from any of the laser-excited states, are shown as dash-dotted lines and in shadowed gray text. (The "5" in the $3D_{5/2}$ state is red to distinguish that atomic state from the [slightly higher-energy] $3D_{3/2}$ state; and the "detuning parameter" $\delta$ is approximately 3.9 GHz.) The dotted black horizontal line in the upper right diagram represents the off-resonant energy corresponding to the frequency of the first (i.e., the yellow-orange) laser in the dual-laser approach. As is shown in these upper two diagrams, both of these LPRS approaches result in "fully-mandated cascades" from the 820 nm (or 819 nm) de-excitation to the 589 nm (or 590 nm) de-excitation, resulting in a mandated 1:1 ratio between those produced photons.

Using calibrations provided by these approaches, the lower two plots show the expected constraints on the cosmological dark energy equation of state parameters $w_0$ and $w_a$, obtained using simulated catalogs of type Ia supernovae corresponding to the expected first three years of observation at the Vera C. Rubin Observatory. In either approach for generating an LPRS, the LPRS results in large expected improvements in the observational constraints on the dark energy cosmological parameters $w_0$ and $w_a$, with the greater of the two expected improvements being from the dual-laser LPRS approach [14].

measured at wavelengths of 589 nm, vs. astronomical magnitudes measured at wavelengths of 820 nm, to be performed at up to 100-fold better precision than the present approximately 1% uncertainties on such measured SNe Ia magnitude ratios [14]. Such an LPRS would, thus, allow for unprecedented precision on future measurements of dark energy.

Figure 2 shows two different approaches that can be used to create such an LPRS (located at an observatory, for example the Rubin Observatory in Chile). In the single-laser approach, a powerful near-UV laser tuned near 343 nm to excite the $3D_{3/2}$ sodium state would be aimed at the sky above an observatory, and would create a cascade of 819/820 nm photons produced in a 1:1 ratio with 589/590 nm photons in the upper atmosphere. And, in the alternative, dual-laser approach, two co-aligned lasers, one near 589 nm and a second near 820 nm, would work together to excite the $3D_{5/2}$ sodium state, which would then de-excite in a cascade of 820 nm photons produced in a 1:1 ratio with 589 nm photons in the upper atmosphere. Although the single-laser approach is slightly simpler in concept; the dual-laser approach would be both simpler to construct in practice, and also would provide a far brighter LPRS that would result in greater improvement on measurements of dark energy than would the single-laser approach. The dual-laser approach requires two lasers instead of one; however, the relatively larger cross-section of the resonances involved in the dual-laser approach means that one can produce a dual-laser LPRS that is over 1000× brighter than that from the single-laser approach, while using lasers that each require only about 5% of the optical output power of the laser that would be required if using the single-laser approach [14].

An LPRS would precisely calibrate results from a ground-based observatory, and thus would not *directly* calibrate space observatories such as JWST; however, by using an LPRS at its ground-based observatory to precisely calibrate a set of stable white dwarf stars, one could then use that set of white dwarf stars to (indirectly, but still precisely) calibrate JWST and other space telescopes — as well as other, separate, ground-based observatories [13b,c].

In the dual-laser LPRS approach, the repeated pulses from the two lasers would be timed such that the STIRAP (STImulated Raman Adiabatic Passage) [15] technique for the excitation of the upper-atmospheric sodium atoms would be implemented. STIRAP is a multi-laser technique that has been commonly used within physical chemistry laboratories around the world since the early 1990s [16], however the STIRAP technique has not yet been utilized in the open atmosphere. An implementation of a two-laser LPRS may thus mark the first utilization and observation of "STIRAP in the sky" — in addition to usage of an LPRS for calibration in cosmology and for the understanding of dark energy, as well as for atmospheric physics and chemistry.

Dark energy is, at present, consistent with being the "cosmological constant" from Einstein's equations of general relativity; and it has been reasonably consistent with being a cosmological constant ever since its initial discovery [1,2,5]. However, amongst other problems [17], this value of a cosmological constant is both unexplained, and unexpected, by the effective quantum field theory that is the Standard Model of particle physics [18]. Also, its relation, if any, to the vastly larger cosmological constant-like expansion that appears to have occurred within the first $10^{-33}$ s after the Big Bang, known as "cosmological inflation," remains unexplained [19]. So, is dark energy a cosmological constant, or not…? It may take some human-generated light — precise artificial stars — in order to determine the true nature of our Universe's dark side.

**FUNDING**

J.E.A. gratefully acknowledges support from Canadian Space Agency grants 19FAVICA28 and 17CCPVIC19. D.B. gratefully acknowledges support from the PRISMA+ Cluster of Excellence funded by the German Research Foundation (DFG); from the European Research Council (project Dark-OST); and from the DFG Reinhart Koselleck project. H.R.S. gratefully acknowledges support from the National Science Foundation grant for the Institute for Theoretical Atomic, Molecular, and Optical Physics (ITAMP) at Harvard University and the Smithsonian Astrophysical Observatory.

Nat. Sci. **2,** e20220003

**PEER REVIEW**

The peer review history for this article is available at https://publons.com/publon/10.1002/ntls.20220003.

This Research Highlight underwent internal review by Gerard Meijer and Bretislav Friedrich.

**ORCID**

*Justin E. Albert*  https://orcid.org/0000-0003-0253-2505
*Dmitry Budker*  https://orcid.org/0000-0002-7356-4814
*H. R. Sadeghpour*  https://orcid.org/0000-0001-5707-8675

[1] A. G. Riess, et al., *Astron. J.* **1998**, *116*, 1009.
[2] S. Perlmutter, et al., *Astrophys. J.* **1999**, *517*, 565.
[3] J. P. Gardner, et al., *Space Sci. Rev.* **2006**, *123*, 485; and https://jwst.nasa.gov.
[4] Ž. Ivezić, et al., *Astrophys. J.* **2019**, *873*, 111; and https://www.lsst.org.
[5] a) D. O. Jones, et al., *Astrophys. J.* **2018**, *857*, 51; b) M. Betoule, et al., *Astron. Astrophys.* **2014**, *568*, A22; c) W. M. Wood-Vasey, et al., *Astrophys. J.* **2007**, *666*, 694.
[6] For projects to provide upcoming improvement on the precision of measurements of *absolute* magnitude values of astronomical sources, rather than on the precision of measurements of *ratios* of such magnitudes as functions of color, please additionally see, for example:
  *i* ) ALTAIR Collaboration 2022 (http://projectaltair.org);
  *ii* ) ORCASat Collaboration 2022 (https://orcasat.ca); and,
  *iii* ) ORCAS Project 2022 (https://asd.gsfc.nasa.gov/orcas/).
[7] C. W. Stubbs, J. L. Tonry, *Astrophys. J.* **2006**, *646*, 1436.
[8] Image from L. Michaille, J. C. Quartel, J. C. Dainty, N. J. Wooder, R. W. Wilson, T. Gregory, *ING Newsletter* **2000**, *2*, 1.
[9] Please see, for example: a) S. S. Olivier, C. E. Max, in *Laser Guide Star Adaptive Optics: Present and Future*, IAU Symp. 158 (Eds: J. G. Robertson, W. J. Tango), Kluwer Press, Dordrecht, Germany **1994**, p. 283; b) D. Bonaccini, et al., in *Adaptive Optics Systems and Technology II*, Proc. SPIE 4494 (Eds: R. K. Tyson, D. Bonaccini, M. C. Roggemann), SPIE Digital Library, Bellingham, WA, USA **2002**, p. 276; c) P. L. Wizinowich, et al., *Publ. Astron. Soc. Pac.* **2006**, *118*, 297.
[10] a) V. M. Slipher, *Popular Astron.* **1929**, *37*, 327; b) S. Chapman, *Astrophys. J.* **1939**, *90*, 309.
[11] J. M. C. Plane, W. Feng, E. C. M. Dawkins, *Chem. Rev.* **2015**, *115*, 4497.
[12] a) P. Juncar, J. Pinard, J. Harmon, A. Chartier, *Metrologia* **1981**, *17*, 77; b) D. E. Kelleher, L. I. Podobedova, *J. Phys. Chem. Ref. Data* **2008**, *37*, 267; c) A. Kramida, Y. Ralchenko, J. Reader, et al., Technical report, *Atomic Spectra Database* (ver. 5.8), NIST **2020**, available at: https://physics.nist.gov/asd.
[13] Please see, for example: a) J. Albert, *Astron. J.* **2012**, *143*, 8; b) J. B. Holberg, P. Bergeron, *Astrophys. J.* **2006**, *132*, 1221; c) R. C. Bohlin, R. L. Gilliland, *Astron. J.* **2004**, *127*, 350.
[14] a) J. E. Albert, D. Budker, K. Chance, I. E. Gordon, F. Pedreros Bustos, M. Pospelov, S. M. Rochester, H. R. Sadeghpour, *Mon. Not. R. Astron. Soc.* **2021**, *508*, 4399 [arXiv:2001.10958]; b) J. E. Albert, D. Budker, K. Chance, I. E. Gordon, F. Pedreros Bustos, M. Pospelov, S. M. Rochester, H. R. Sadeghpour, *Mon. Not. R. Astron. Soc.* **2021**, *508*, 4412 [arXiv:2010.08683].
[15] U. Gaubatz, P. Rudecki, S. Schiemann, K. Bergmann, *J. Chem. Phys.* **1990**, *92*, 5363.
[16] a) K. Bergmann, N. V. Vitanov, B. W. Shore, *J. Chem. Phys.* **2015**, *142*, 170901; b) N. V. Vitanov, A. A. Rangelov, B. W. Shore, K. Bergmann, *Rev. Mod. Phys.* **2017**, *89*, 015006.
[17] Please see, for example: M. Turner, *Phys. Today* **2003**, *56*, 10.
[18] a) S. Weinberg, *Rev. Mod. Phys.* **1989**, *61*, 1; b) P. J. E. Peebles, B. Ratra, *Rev. Mod Phys.* **2003**, *75*, 559.
[19] Please see, for example: C.-I. Chiang, J. M. Leedom, H. Murayama, *Phys. Rev. D* **2019**, *100*, 043505.